\documentclass{ws-jai}
\usepackage[flushleft]{threeparttable}
\usepackage{graphicx}
\usepackage{graphics}
\usepackage{color}
\usepackage{hyperref}
\usepackage{verbatim}
\usepackage{xfrac} % needed for \sfrac{}{}
\usepackage{xcolor}

\usepackage{dcolumn}% Align table columns on decimal point
\usepackage{bm}% bold math
\usepackage{lineno}
\graphicspath{{figures/}{figures2/}}
\DeclareGraphicsExtensions{.pdf,.gif,.jpg,.png}
\begin{document}

%%%%%%%%%%%%%%%%%%%%%%%
% Define new commands %
%%%%%%%%%%%%%%%%%%%%%%%
\newcommand{\rtHz}{$\sqrt{\mbox{Hz}}$}
\newcommand{\pArtHz}{$\frac{\mathrm{pA}}{\sqrt{\mathrm{Hz}}}$}
\newcommand{\phinot}{\mbox{$\Phi_0$}}
\newcommand{\degree}{\mbox{$^{\circ}$}}
\newcommand{\fortran}{{\tt Fortran~77}}
\newcommand{\CXX}{C++}
\newcommand{\order}{\mbox{${\cal O}$}}
\newcommand{\loopgain}{\mbox{${\cal L}$}}
\newcommand{\const}{\mbox{\sc\small Const}}
\newcommand{\approxlt}{ \stackrel{<}{\sim} }
\newcommand{\approxgt}{ \stackrel{>}{\sim} }

\newcommand{\mycomment}[1]{ }
\newcommand{\note}[1]{\mycomment{#1}}
\newcommand{\hilight}[1]{#1}

\newcommand{\partno}[2]{{ #1 (#2)}}
    % e.g. \partno{THS4131}{Texas Instruments, address}

\newcommand\arcdeg{\mbox{$^\circ$}}%
\newcommand\arcmin{\mbox{$^\prime$}}%
\newcommand\arcsec{\mbox{$^{\prime\prime}$}}%
\newcommand\onehalf{\mbox{$\sfrac{1}{2}$}}%
\newcommand\onethird{\mbox{$\sfrac{1}{3}$}}%
\newcommand\twothirds{\mbox{$\sfrac{2}{3}$}}%
\newcommand\onequarter{\mbox{$\sfrac{1}{4}$}}%
\newcommand\threequarters{\mbox{$\sfrac{3}{4}$}}%

\def\apjl{ApJ}
\def\apj{ApJ}
\def\aj{AJ}
\def\mnras{Mon. Not. R. Astron. Soc.}
\def\ApJ{ Astrophys. J.}
\def\ApJS{ Astrophys. J. Supp.}
\def\apjs{ Astrophys. J. Supp.}
\def\AJ{ Astronom. J.}
\def\PrASA{ Proc. Astron. Soc. Aust.}
\def\PIEEE{ Proc. IEEE}
\def\IEEEAPSN{ IEEE Ant. and Prop. Soc. Newsletter}
\def\CS{ Current Sci.}
\def\NPS{ Nat. Phys. Sci. }
\def\prd{Phys. Rev. D}
\def\nat{ Nature}
\def\araa{ Annual Review of Astronomy and Astrophysics }
\def\nar{New Astronomy Reviews}

%--------------------------------------------------------------
%\preprint{APS/123-QED}

\title{ICE-based Custom Full-Mesh Network for the CHIME High Bandwidth Radio Astronomy Correlator}

\newcommand{\mcgill}{\dagger}
\newcommand{\wvu}{\ddagger}
\newcommand{\cifar}{\S}
\newcommand{\threespeed}{\star}

\author{ K. Bandura$^{\wvu,\mcgill}$, J.F.\ Cliche$^\mcgill$, M.A.\ Dobbs$^{\mcgill,\cifar}$, \\
A.J.\ Gilbert$^\mcgill$, D. Ittah$^{\mcgill}$, J.\ Mena~Parra$^\mcgill$, G.\ Smecher$^{\mcgill, \threespeed}$ }

\address{
$^\mcgill$Physics Department, McGill University, Montreal, Quebec H3A~2T8, Canada \\
$^\cifar$Canadian Institute for Advanced Research, Toronto, Canada \\
$^\wvu$LCSEE, West Virginia University, Morgantown, WV 26505 \\
$^\threespeed$ Three-Speed Logic, Inc., Vancouver, Canada \\
}

\maketitle
\corres{$^\S$Kevin Bandura}

\begin{history}
\received{(to be inserted by publisher)};
\revised{(to be inserted by publisher)};
\accepted{(to be inserted by publisher)};
\end{history}

\begin{abstract}

  New generation radio interferometers encode signals from thousands of antenna feeds across
  large bandwidth. Channelizing and correlating this data
  requires networking capabilities that can handle unprecedented
  data rates with reasonable cost. The Canadian Hydrogen Intensity
  Mapping Experiment (CHIME) correlator processes 8-bits from $N=2048$
  digitizer inputs across 400~MHz of bandwidth. Measured in $N^2~\times
  $ bandwidth, it is the largest radio correlator that has been built.
  Its digital back-end must exchange and reorganize  the 6.6~terabit/s produced by its 128 digitizing and
  channelizing nodes, and feed it to the 256-node spatial correlator in a way
  that each node obtains data from all digitizer inputs but across a small
  fraction of the bandwidth \hilight{(i.e.
  `corner-turn').} In order to maximize performance and
  reliability of the corner-turn system while minimizing cost, a
  custom networking solution has been implemented. The system
  makes use of Field Programmable Gate Array (FPGA) transceivers to implement direct, passive, full-mesh, high speed serial
  connections between sixteen circuit boards in a crate, to
  exchange data between crates, and
  to offload the data to a cluster of 256 graphics
 processing unit (GPU) nodes using standard 10~Gbit/s Ethernet links. The GPU nodes complete the corner-turn by combining data from all crates and then computing visibilities. Eye diagrams and frame error counters confirm error-free operation of the corner-turn network in both the currently operating CHIME Pathfinder telescope (a prototype for the full CHIME telescope) and a representative fraction of the full CHIME hardware providing an end-to-end system validation.
  An analysis of an equivalent corner-turn system built with Ethernet switches instead of custom passive data links is provided.

\end{abstract}


%--------------------------------------------------------------

%!TEX root = ICE-CornerTurn_main.tex

\section{Introduction}
\label{s_introduction}

A new generation of radio telescopes operating or in development, including
LOFAR \citep{2009IEEEP..97.1431D},
MWA \citep{2009IEEEP..97.1497L},
LWA \citep{2009IEEEP..97.1421E},
UTMOST \citep{2016MNRAS.458..718C},
CHIME \citep{2014SPIE.9145E..22B},
HERA \citep{2014ApJ...782...66P},
\hilight{HIRAX \citep{2016arXiv160702059N},}
and SKA \citep{2004NewAR..48..979C},
employ arrays
with hundreds or thousands of antenna feeds and large bandwidth, as opposed to
more traditional radio interferometers with fewer feeds and large dishes.
This moves cost and key challenges of observatory
design and construction away from the mechanical structure and cryogenic
receivers to digital signal
processing, where the information from the feeds is digitized and
processed in a radio correlator.

Creating a large, distributed frequency-spatial (FX) correlator system that can process this much data poses
a networking challenge \citep[see e.g.][]{2011MNRAS.410.2075L}.
Indeed, the digitized sky information acquired and channelized
by each acquisition node (F-engine) describes the signal content of a few antenna feeds over the whole frequency band.
However, each
spatial correlation node (X-engine) requires the data from all antenna feeds, but for a subset of the frequency
band, so
that all inputs can be multiplied together and integrated to form visibilities. This requires an exchange of data that is called the corner-turn, which is akin to a matrix transpose.
The corner-turn process implements this transpose,
going from a configuration where all the information from one antenna
feed across all frequency channels is in one location at the F-engine, to a
configuration where all the information from one frequency channel
across all antenna feeds is in one location at the X-engine.
The
total data rate is given by the product of
the frequency bandwidth multiplied by the number of inputs
\begin{equation}
    \mbox{Data\ Rate} = N \ \times \   BW \ \times \ R_\mathrm{bits} ,
\end{equation}
where $N$ is the number of digitizer inputs (twice the number of dual polarization antennas), $BW$ is
the operating frequency bandwidth, and $R_\mathrm{bits}$ is the number of
bits per frequency channel.  Note that the data-rate is independent of the
number of frequency channels that have been produced by the
F-engine.

The Canadian Hydrogen Intensity Mapping Experiment
(CHIME)
 instrument is composed of four
20$\times$100~m cylindrical reflectors instrumented with 1024 dual
polarization feeds. Its FX-architecture correlator processes data
from $N$=2048 FPGA-based digitizer inputs across $BW$=400~MHz of bandwidth with $R_\mathrm{bits}$=8 bits per
frequency channel (4~bits real and 4~bits
imaginary), resulting in a total of 6.6~Tbit/s of data that needs to
be reorganized by a corner-turn system before being fed into
256 GPU-Based correlator nodes that compute the full $N^2$ correlation
matrix.

The CHIME GPU X-engine is described in Refs.~\citet{Denman:2015ec} and \citet{Recnik:2015ev}.
A GPU node refers to a networked host with a $4~\times$ 10~Gbit/s network card and
multiple GPU cards. For CHIME, the nodes have been sized such that the volume of data produced by
the FPGA boards may be processed, as well as enabling additional on-site
experiments to have real-time data access with some pre-processing.
At minimum, each GPU node in the X-engine would process one frequency channel.
Through careful software and hardware design \hilight{each node in this system processes four frequency
channels.}

\hilight{The hybrid nature of the correlator system (FPGA based F-engine, GPU based
X-engine) was driven by the ability for rapid development and flexibility.  The
GPU based X-engine enables the flexibility to explore new real-time calibration
and expand the capabilities of the system.  The hybrid design has already
allowed for the addition of a Pulsar back-end to the system, enabling the search
and monitoring of pulsar-like objects.  }

Overall, when CHIME comes online this year, it will be the
largest radio correlator that has been built, measured in number of digitizer
inputs squared times bandwidth ($N^2\times BW = 1.7 \times 10^{15}$
multiplies per second). Given the massive amount of data that needs to be moved around, an efficient
corner-turn architecture is a cost and complexity driver for the CHIME correlator.

This
paper describes the CHIME corner-turn implementation with FPGA-based hardware and firmware.
Section~\ref{s_shuffle} describes the corner-turn system architecture, while Section~\ref{s_design} describes the hardware needed to implement it.  In Section~\ref{s_performance} we describe the testing and
performance of the system and Section~\ref{s_comparisons} compares the CHIME
corner-turn system \hilight{with another candidate architecture.}

%--------------------------------------------------------------
%!TEX root = ICE-CornerTurn_main.tex

\begin{figure}[htbp]
\centering
\includegraphics[width=.9\textwidth]{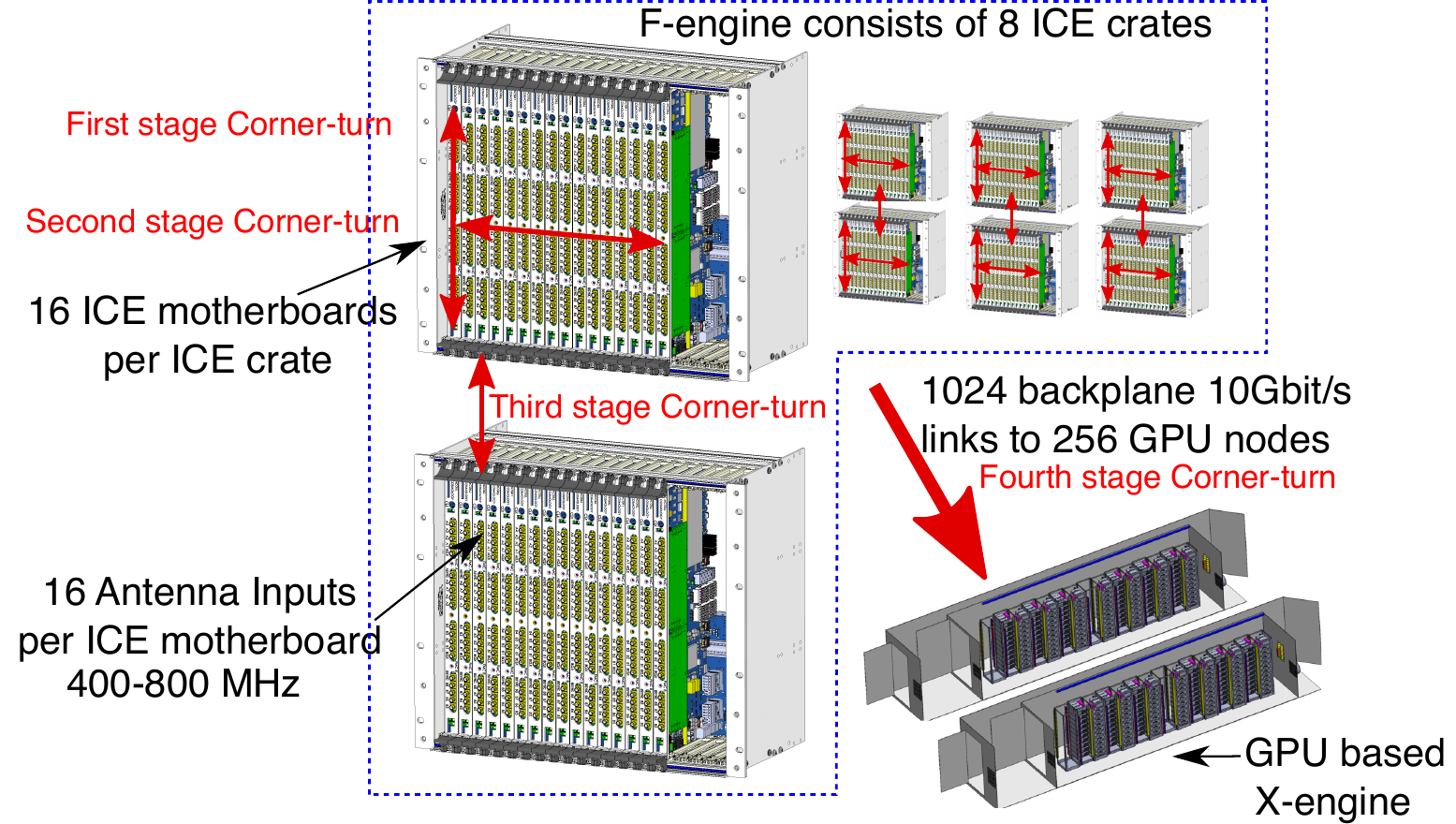}
\caption{
Block diagram of the overall architecture of the CHIME digital back-end. Eight
crates of 16 boards digitize 2048 analog signals. The corner-turn operation
is performed in multiple stages. First, a data corner-turn is performed within a
single ICE motherboard by reordering all channelized antenna data by channel number
rather than antenna. Second, boards within a crate exchange data through the 3.8~Gbit/s (10~Gbit/s capable) full-mesh passive backplane network.
A third corner is done through the 40 passive 7.6~Gbit/s
copper links between pair of crates. The data from all boards is then sent to
the remote GPU array over 256, quad 10~Gbit/s optical cables, where each GPU
merges the data from the ICE motherboards in four pairs of crates to complete the 
corner-turn operation.
}
\label{FigDataFlow}
\end{figure}

\section{Corner-turn System Architecture}
\label{s_shuffle}

The CHIME custom corner-turn network was implemented using the `ICE'\footnote{ICE is the name of the system, and is not an acronym.} electronics framework~\citep{2016ICE.SYSTEM.PAPER}.

The decision to develop custom hardware and firmware instead of using an array of 
Ethernet
switches (an alternative described in Section~\ref{s_comparisons}) 
was based on a combination of cost, performance, technical risks and schedule
risks.
Since the FPGA required to perform the data acquisition and channelizing provides
plenty of multi-gigabit transceivers capable of transmitting data
over passive links, and since a backplane was already required to
distribute power and timing, interconnecting the ICE motherboards to perform the corner-turn was a matter
of adding copper traces and connectors to the backplane--a negligible additional cost for the system. 
In addition to the performance and risk drivers that pushed the design towards the custom solution, the cost of an array of switches with an instantaneous,
non-blocking bandwidth of \hilight{6.6}~Tbit/s would have been significant
in comparison to the total cost of the CHIME F-engine (about \$2M including hardware and design engineering).

The overall architecture of the CHIME digital back-end is presented in
Figure~\ref{FigDataFlow}. Sky-signals are digitized and channelized at the 2048
inputs of the F-engine which are distributed across 128 custom FPGA-based ICE
motherboards~\citep{2016ICE.SYSTEM.PAPER} housed in eight \hilight{standard 19" subracks (or
crates)}. Each
channelizer partitions the 400~MHz bandwidth data into 1024 frequency channels,
each truncated to a complex number consisting of 4~bits real and 4~bits imaginary. 
The corner-turn system needs to reorder the
data such that the 256 GPU correlator nodes each end up with a unique set of
$1024/256=4$ frequency channels from every one of the 2048 channelizers.

\begin{figure}[htbp]
\centering
\includegraphics[width=1\textwidth]{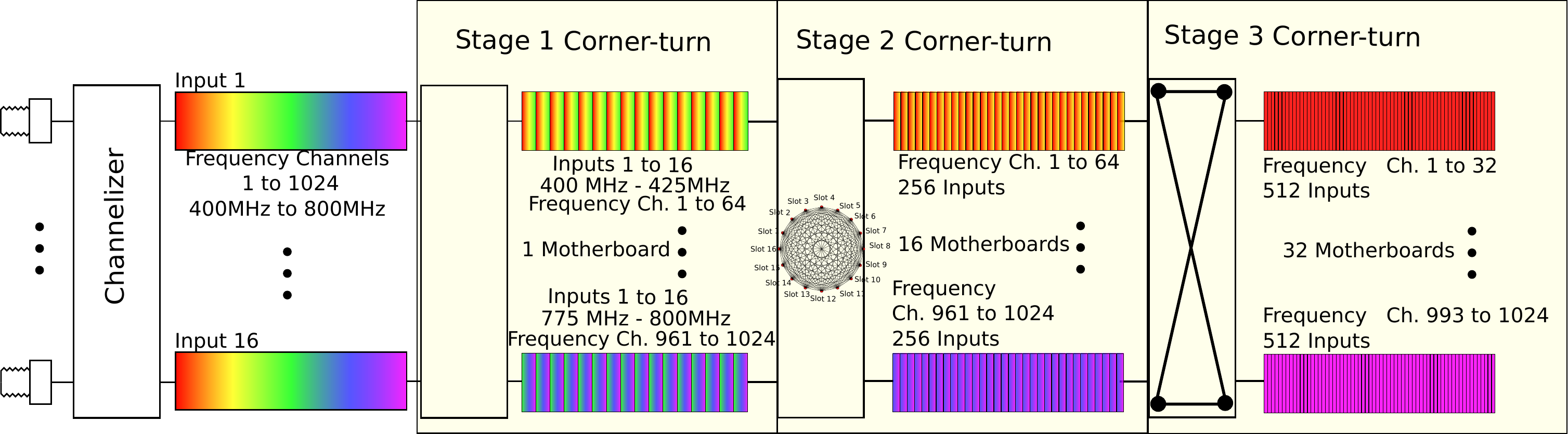}
\caption{Diagram showing the data flow from the digitizer input through \hilight{to the third} stage
corner-turn. Data from 16 digitizer inputs arrive at each ICE motherboard and is
channelized into 1024 frequency channels. In stage one, this data is re-organized into 16
different packets each containing the information from all 16 digitizers across
64 frequency channels. In the stage two corner-turn, each motherboard in a crate
is designated 1/16th of the frequency spectrum. \hilight{ Using the full mesh
communication, each motherboard receives 64 frequency channels from all other
boards in the crate and transports all other channel data to their designated
locations. Once this is complete, the stage three corner-turn takes place through
the backplane QSFP+ connections, and half the frequency channels are provided to a sister 
motherboard in a
second crate. The net result is that each motherboard has data for 512 inputs for 32
frequency channels.}}
\label{FigShuffleStage123}
\end{figure}

\begin{figure}[htbp]
\centering
\includegraphics[width=.5\textwidth]{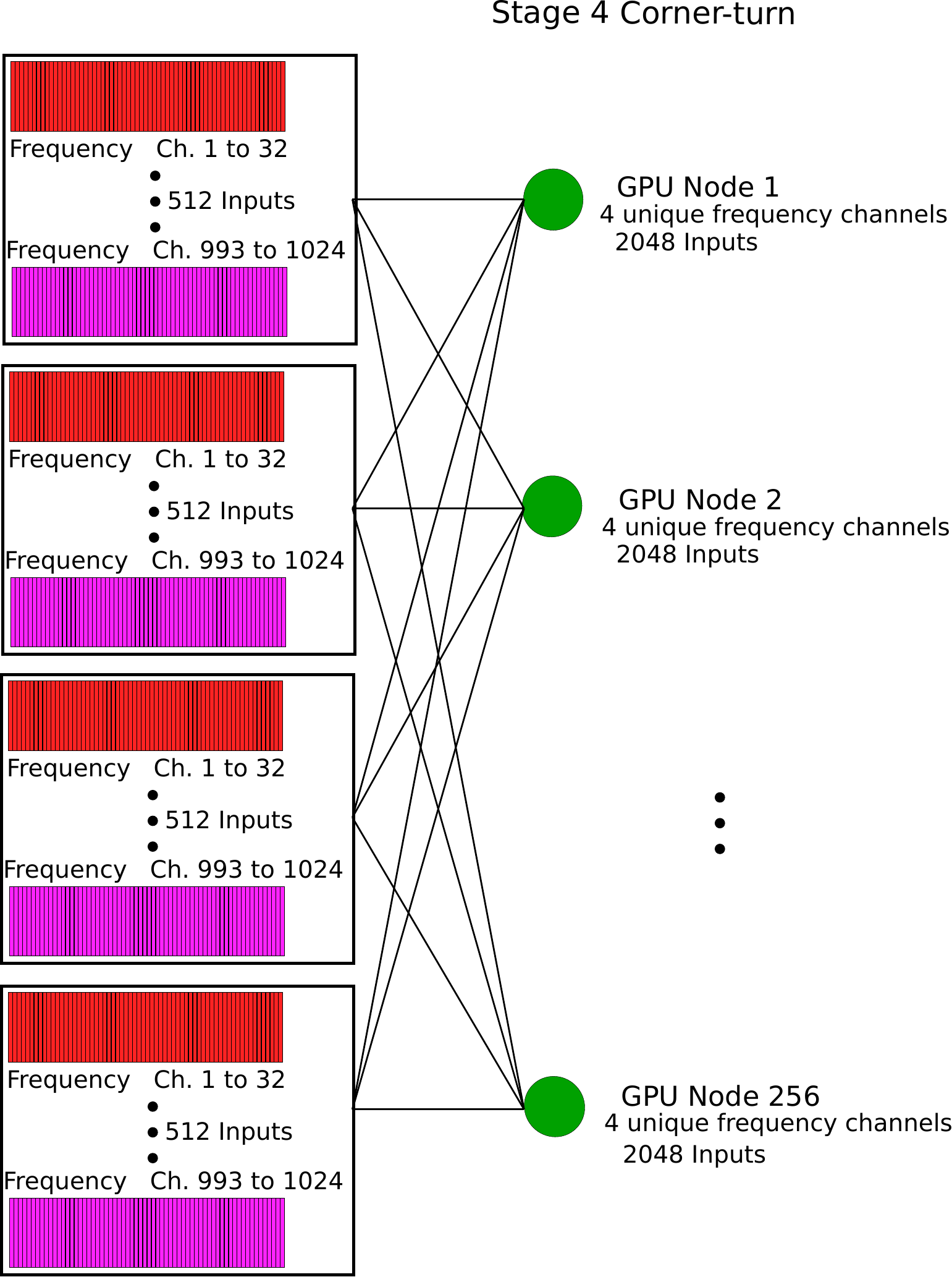}
\caption{Diagram showing the data flow through the stage four
  corner-turn \hilight{ from the ICE motherboards to the GPU nodes.
  Each ICE motherboard has all the data from 512 inputs across 32
  frequency channels and is connected through eight 10~Gbit/s Ethernet ports to
  eight different GPU nodes. Each GPU node receives from four different ICE 
  motherboards, such that every GPU node ends up with data from all 2048
  digitizer inputs for four unique frequency channels. } }
\label{FigShuffleStage4}
\end{figure}

Overall, the corner-turn operation described above is performed completely by the
ICE FPGA-based motherboards and cables in four stages as illustrated in
Figures~\ref{FigShuffleStage123} and \ref{FigShuffleStage4} and described below.
\hilight{All four stages are implemented using a single VHDL module that is configurable for each stage.}
\hilight{Note that the bit rates in this section correspond to data rates attributed to the raw data that \hilight{are} being transmitted. Actual line rates are higher, because the packets that are transmitted also include packet headers, error checking cookies, and flagging information.
The -2 speed grade Xilinx Kintex 7 series FPGAs used for the ICE system support line rates of 10.2~Gbit/s. Faster speed grades are available that support line rates up to 12.5~Gbit/s.}

The first stage is implemented within each FPGA, with no data exchanged between motherboards.  
Each ICE motherboard processes
sixteen analog inputs (51.2~Gbit/s raw data rate). The FPGA receives the 16 inputs and creates
16 new data streams, each of which combines 1/16th of the frequency channels from
each of the 16 channelizers.

During the second stage, 15 of these data stream are sent to the other boards
within a crate through the backplane mesh network, with one stream remaining
local. The end result is that each ICE motherboard in a crate slot gathers data for a
specific subset of frequency bins (64 channels total), but from all 256
channelizers in the crate. Approximately 4~Gbit/s of \hilight{raw} data, flags and headers are
sent by each backplane link, resulting in about 1~Tbit/s of backplane full-mesh
traffic.  The packets received through the 15 backplane
links are reordered into two streams, with each stream containing half of the
received frequency bins.

There are several supported configurations for the third stage, see Section~\ref{subsec_shuffle}.
For the CHIME implementation,  one of the new streams created in stage two is sent to the
ICE motherboard at the same slot position in the adjacent crate through four 10 Gbit/s
links carried by the backplane Quad Small Form-factor Pluggable Plus (QSFP+)
connector. About 7.4~Gbit/s \hilight{raw data and flags are} sent over each of the four links through commercial passive 3\ m copper QSFP+ cables. 
Similarly the
other crate provides one of its two streams over the same QSFP+ link. In the
end, each ICE motherboard within the pair of crates ends up with a unique subset of 32
frequency bins from 512 channelizers. For CHIME this corresponds to the data
from one quarter of the array, or one cylinder.  The packets from the local and
remote stream are reordered again into eight new streams containing data from
512 analog inputs and four frequency channels.

In the fourth stage, the eight streams from the third stage are provided to
eight separate GPU nodes using $8~\times$ 10~Gbit/s Ethernet links provided by the two
QSFP+ connectors located on the ICE motherboard. Each Ethernet link carries about
7.5~Gbit/s \hilight{raw data and flag rate}. Two active 100\ m multi-mode optical fiber QSFP+ cables, each terminated into
four independent Small Form-factor Pluggable Plus (SFP+) connectors, are used to
carry the data from the ICE motherboards to the GPU nodes, which are located in different buildings. Each SFP+ connector
from an ICE motherboard is connected to a different GPU node. A single
four-port 10~Gbit/s Ethernet card at each GPU node accommodates the
direct links. Overall, a GPU node receives data from a FPGA on an ICE
motherboard in each of the four quadrants, and therefore possesses data from four
unique frequency bins from all 2048 channelizers.

The final corner-turn step is performed by the host processor on the GPU node 
\citep{Recnik:2015ev,Denman:2015ec,Klages:2015em},
which recombines the data from its four 10~Gbit/s Ethernet ports to generate
four contiguous data blocks containing the data for one frequency bin from every
channelizer. These data blocks can then be sent efficiently over the PCI Express
bus to the four GPU cores located in the node, where they are correlated.

%--------------------------------------------------------------
%!TEX root = ICE-CornerTurn_main.tex

\section{Hardware Implementation}
\label{s_design}

\begin{figure}[htbp]
\centering
\includegraphics[width=.60\textwidth]{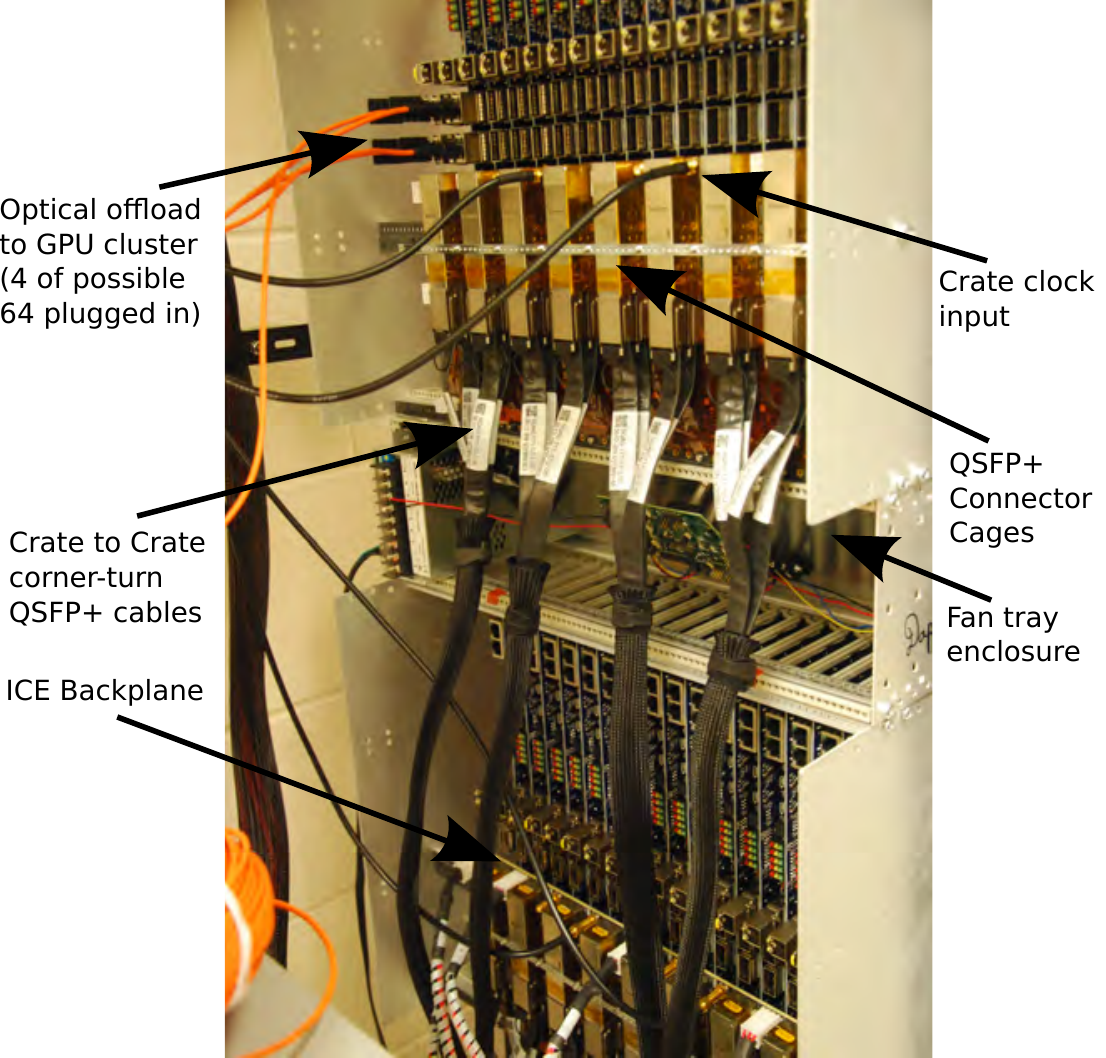}
\includegraphics[width=.38\textwidth]{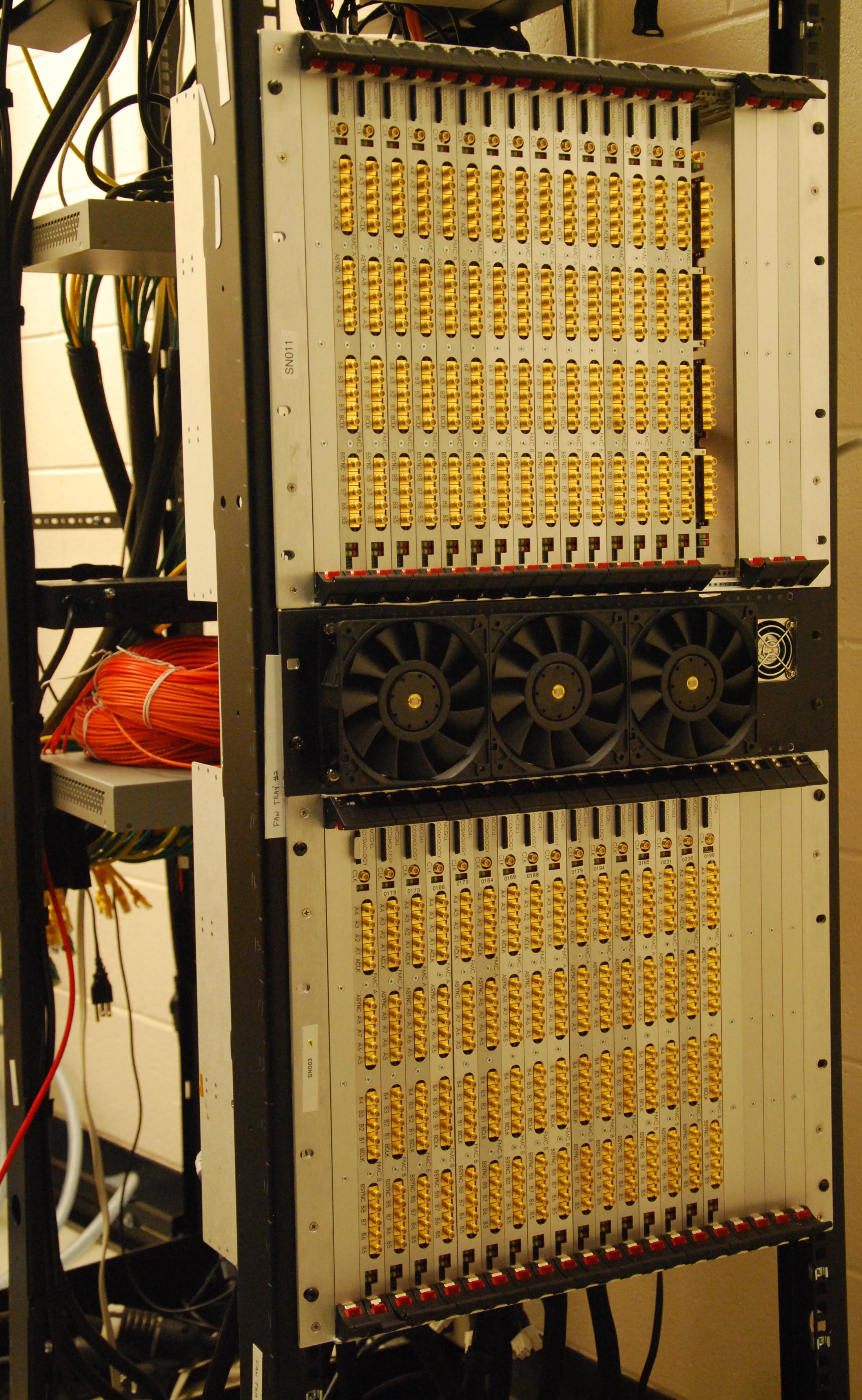}
\caption{Rear and front view of a pair of crates that form one quarter of the
CHIME F-engine and corner-turn system. The rear view shows the crate-crate
corner-turn cabling as well as a few of the optical connections from the
motherboard to GPU nodes (for clarity, only 4 of the 64 QSFP+ cables that are
installed). The front view shows the high-density analog inputs and the crate
water-based cooling system (the radiator is not installed in this photo).}
\label{FigBackplanePhotoShuffle}
\end{figure}

Physically, the corner-turn system \hilight{is implemented on} the same hardware as the F-engine
and is distributed across the 128 ICE motherboards that constitute the CHIME
correlator F-engine.

Each ICE motherboard hosts two custom full-size FPGA Mezzanine cards (FMC, following the ANSI/VITA 57 standard) which
digitize eight analog inputs each. This configuration yields a reasonable board
height (9U), a high channel density and requires a number of inputs/outputs (I/O) that is
compatible with \hilight{lower-cost FPGA packages}. In addition to its backplane
connectivity, each board offers two QSFP+ links that can be
used by the FPGA to implement eight bidirectional links at up to 10 Gbit/s each.

Sixteen ICE motherboards are packaged densely into a \hilight{crate} and are connected through a custom backplane which allows for power and
timing signal distribution to all boards in the crate. The backplane also
provides links between boards located in the same crate and other crates.
Figure~\ref{FigBackplanePhotoShuffle} shows a photo of two interconnected
crates, which constitute one quarter of the CHIME's F-engine and corner-turn
system.

The CHIME F-engine and corner-turn system consists of eight crates of ICE motherboards
housed in two RF-shielded buildings located below the telescope structure. The data is
sent with optical 10~Gbit/s Ethernet links about 100~m away to two separate RF-shielded buildings housing the X-engine GPU nodes.

A global positioning system (GPS) receiver, equipped with an integrated clock
distribution module, generates a 10~MHz reference clock and
Inter-Range Instrumentation Group B (IRIG-B, Standard 200-04)
timestamps that are distributed to each crate via coaxial cable. The
backplane distributes these to each of the motherboards.  The 10 MHz
reference clock is also used to generate all the FPGA clocks and ensures
synchronized data acquisition, framing and transmission throughout the array.

The data links provided by the motherboards and backplane are the backbone of
the corner-turn system and are described below.
Information on the rest of the ICE hardware can be found in Ref.~\citet{2016ICE.SYSTEM.PAPER}.

\subsection{Full-mesh backplane corner-turn hardware}
\label{subsec_full_mesh}

Stage two of the corner-turn utilizes a full-mesh duplex network that
connects every board  with every other board in a sixteen-slot ICE crate with
a bidirectional link designed to operate at up to 10~Gbit/s in each direction.
This network consists of 240 bidirectional high speed serial links embedded in
the backplane circuit board. CHIME uses about 40\% of the  2.4~Tbit/s
theoretical capacity of that network.

Maintaining proper signal integrity of 10~Gbit/s links routed on circuit
boards and through connectors requires great care. In order to reduce cross-talk,
attenuation and reflections as much as possible, every link consists of
differential copper lines implemented as edge-coupled 100$\Omega$ stripline
routed on a single layer from the source to the destination point.
25~Gbit/s-rated Molex Impact connectors were used to interface the backplane to
the
motherboards. Signal reflections are further minimized by back-drilling each
of the connector press~fit pads. Ground planes separate each of the eight
signaling layers required to perform the full mesh. \hilight{24} layers were required to
complete the backplane routing\hilight{, with eight signal layers performing the full-mesh}.

Circuit board material generally suffers high losses at 10\ Gbit/s and had to be
chosen carefully to balance cost and performance. Since most of the high speed
signal-length occurs over the backplane, it was constructed from Panasonic
Megtron~6, a very low-loss hydrocarbon resin material with similar performance
to \hilight{even} more expensive hydrocarbon resin materials with ceramic filler (such as
Rogers 4350B) that are commonly used for high-frequency, low-loss applications.
The maximum attenuation at 5\ GHz is expected to be 0.37\ dB/inch at 10\ Gbit/s,
resulting in a maximum backplane attenuation of 6\ dB between the furthest
slots. The material is expensive, but its cost is justified by the overall
signal-integrity advantages and the low number of backplanes in the system.

The motherboards are constructed from  a lower-cost 408HR woven glass FR-4
material. The transmission distance link between the FPGA and backplane connector
results in an additional 5~dB attenuation.

To further reduce losses, both the backplane and motherboard laminates have
been fabricated using very low profile (VLP) copper foil. VLP reduces losses
by 0.2~dB/inch for cost increase of only about 2\%.

Overall, the worst-case FPGA-to-FPGA attenuation is 16~dB for the longest trace.
This is well within the specifications of the Xilinx Kintex 7 GTX
transceivers, which can operate at full rate with greater than 20~dB
attenuation \citep[see][]{XilinxUG476}. In practice, the links operate as
expected. Test results are described in Section~\ref{s_comparisons}.

\subsection{Inter-crate corner-turn hardware}
\label{subsec_shuffle}

\begin{figure}[!htbp]
  \centering
    \raisebox{-0.5\height}{\includegraphics[width=0.48\textwidth]{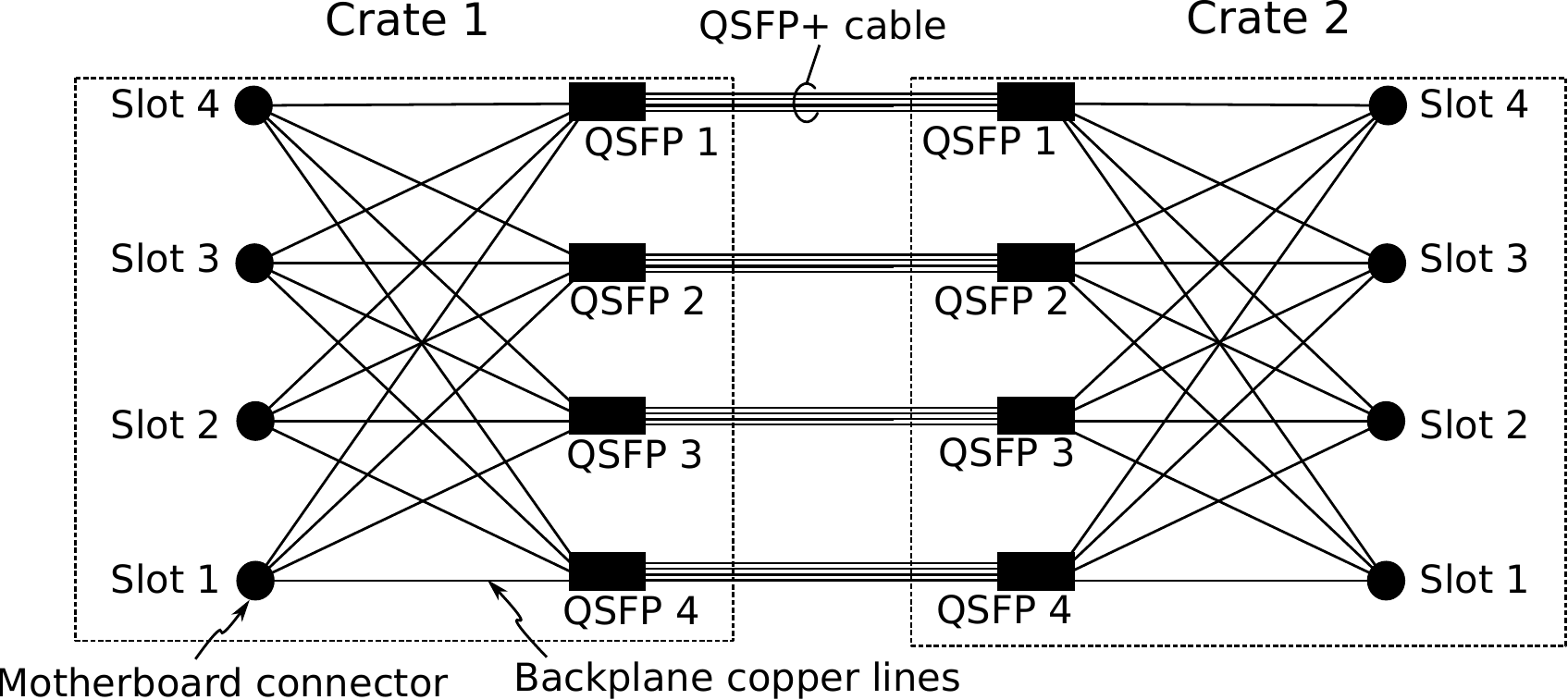}}
    \raisebox{-0.5\height}{\includegraphics[width=0.48\textwidth]{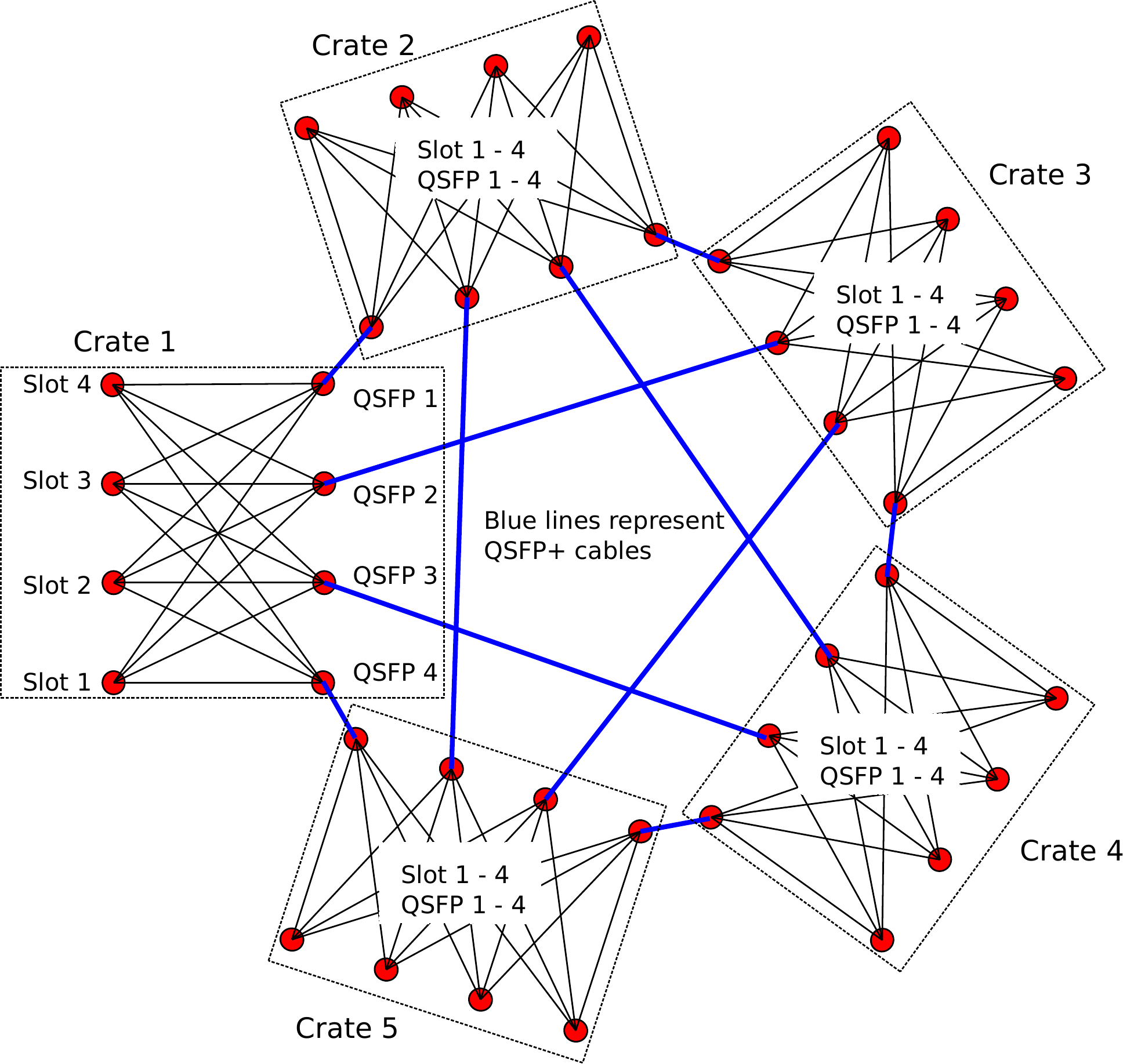}}
    \caption{Diagram showing the connectivity of the QSFP+ ports on the
      ICE backplanes. The four links in each QSFP+ connector connect to four independent ICE motherboards, \hilight{as shown for each crate in the diagram. Two inter-crate stage three configurations are shown.}
      Left: In CHIME, data is exchanged between pair of crates by simply connecting corresponding QSFP+ ports. \hilight{There are 16 such QSFP+ connections per crate.}
       Right: The interleaved connection allows the creation of a full mesh network between up to 5 crates, allowing an even larger array to be built.  }
  \label{FigQSFPMesh2crates}
\end{figure}

Stage three of the corner-turn operation relies on the
transmission of data between boards located on the same slot in multiple crates. To achieve
this, each backplane includes sixteen QSFP+ ports,  each capable of four
bidirectional 10~Gbit/s links. The connections between the motherboard's FPGA and the backplane's QSFP+ ports use the same signal-integrity
control method as those used for the full-mesh intra-crate network described
above. Each motherboard provides four bidirectional links which fan out to four
adjacent QSFP+ ports \hilight{on the backplane} in such a way that each QSFP+ port is linked to four
neighboring motherboards, as shown for each crate in Figure~\ref{FigQSFPMesh2crates}.

Stage three allows the formation of a full-mesh network between
motherboards in up to five crates by using simple off-the-shelf
QSFP+ cables.  A two-, three- or five-crate network respectively provides 40~Gbit/s (four 10~Gbit/s links), 20\ Gbit/s (two 10\ Gbit/s links) or one 10\ Gbit/s link between each motherboard.
The two-crate configuration is shown on the left side of Figure~\ref{FigQSFPMesh2crates} and is used for CHIME.
The five-crate configuration is shown on the right side of Figure~\ref{FigQSFPMesh2crates}. The bandwidth required for
each crate-crate connection is the total amount of channelizer data processed by one motherboard
divided by the number of crates participating in the corner-turn. For CHIME,
each motherboard \hilight{has} \hilight{approximately} 60~Gbit/s (data + overhead) to corner-turn. A two-crate
corner-turn of the current CHIME system therefore requires 30~Gbit/s \hilight{between motherboards} (75\% of
the available bandwidth), whereas a three- or five-crate system would require 20~Gbit/s
or 12~Gbit/s respectively. In the latter case, the bandwidth available is insufficient to transmit all the data with the overhead of packet headers and flags that are presently included. However, by removing several interference-contaminated
frequency channels and reducing redundant data flagging information, the data could be made to fit in the available bandwidth. An alternative solution would be to use a better speed-grade version of the same FPGA model, which accommodates 12.5~Gbit/s serial line rate transmission.

For crates that are in close proximity (up to about 7~m), passive copper QSFP+
cables can be used. For longer inter-crate separations, active copper or
optical transceiver cables may be used. CHIME uses 2~m QSFP+ flat cables
 with excellent results and has tested cables up to 7~m in length.

\subsection{GPU links} 
\label{sec:gpu_links}

Once the FPGA-based corner-turn operation is completed, each motherboard in
the array possesses \hilight{32 frequency channels, broken up into eight} streams of data each containing the channelized data
for \hilight{4} frequency channels.
The motherboards \hilight{send their} data to eight
separate GPU nodes using 10~Git/s Ethernet links provided by two QSFP+ ports
on the motherboards.
The GPU nodes are located in two separate server buildings.
One 56~m cable and one 100~m
cable, each terminated by four SFP+
connectors, connect from ICE motherboards to GPU nodes.
The four-fiber, 3~mm diameter active optical cable uses OM2+ multimode fiber and offers a high single-cable data carrying capacity at a low price. The cable was
experimentally tested to ensure error-free operation.

\subsection{Firmware} 
\label{sub:firmware}

In addition to the data acquisition and channelizing, the FPGAs on the
ICE motherboards are responsible for performing the first three stages of
the corner-turn while using a minimum amount of the FPGA logic resources. \hilight{The custom firmware that has been developed for this purpose is configurable, such that the same code is reused for each of the stages.}

Each stage of the corner-turn operation starts with a frame alignment module
that buffers the frames and ensures they are aligned when passed down the
processing pipeline. This is necessary because the channelizers operate on
multi-phase clocks and the networking links have small but significant
differential delays. Small FIFOs are used to perform the alignment. Since the
transmission of data is \hilight{performed} synchronously across the array, data will always
arrive in the right order and within a very small window of time relative to a
reference link. This allows the alignment module to quickly give up on missing
packets and forward the remaining data (with appropriate flags) without
requiring large memory buffers within the FPGA.

Once the packets are realigned, the reordering modules repackage the incoming
data into a number of output streams generated by channel selector modules
that pick specific frequency channels from all input streams. Programmable
tables determine which channels are gathered within each output stream. Data
from multiple frames can also be combined into single packets to improve
transmission and processing efficiency. \hilight{This is another advantage of the custom corner-turn strategy, which allows larger, more efficient packets to be built up as data from a narrow frequency bandwidth is amalgamated.}

The channel selectors use local buffers at every stage of the corner-turn to group the
data by frequency channel and to forward it as a single block to the next
stage. This will ultimately allow the GPU node to efficiently transfer the
data it receives to the GPU processor in a minimum number of memory transfer operations\hilight{--a crucial feature to reduce the processing overhead that would otherwise complicate the host IO handling dramatically.}

Each frequency sample is tagged with a 1-bit saturation flag. The flags from all
selected channels are also packed into words and follow the data as a single
block that can be quickly scanned by the GPU node to identify bad data. A
third block of flags indicate whether each of the channelizers involved in the
data suffered from overflows during the analog-to-digital conversion or during
the Fourier transform process. \hilight{Since the networking system has extra bandwidth available, the flagging protocol is optimized to simplify the data handling tasks for the GPU host, rather than minimizing the size of the transmitted data.}

Each stage of the corner-turn firmware provides bypass paths such that the firmware can
be configured on-the-fly as a single-board (16 input), single-crate (256
input) or dual-crate (512 inputs) F-engine/corner-turn system. 

\begin{figure}[!htbp]
  \centering
    \includegraphics[width=0.8\textwidth]{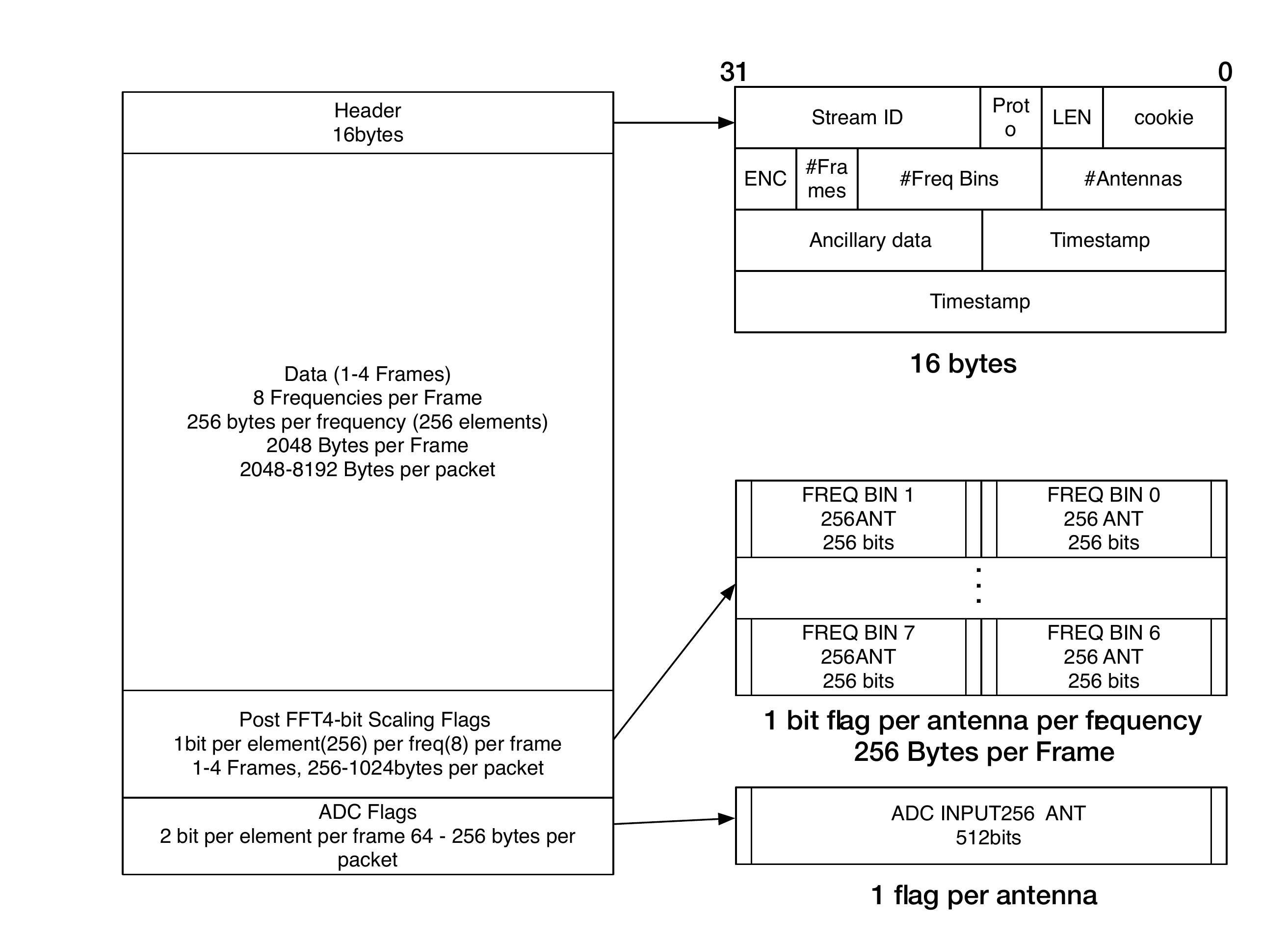}
    \caption{ Illustration of the packet format used to
    carry the information between the various corner-turn stages. Illustrated here
    is the packet after the second corner-turn.  }
  \label{Fig:packet_format}
\end{figure}

Each packet transiting through the backplane mesh network, the inter-crate corner-turn
links, and the GPU links is preceded by a header that indicates the packet
geometry (number of channels and inputs), its origin (stage, crate, slot and
lane), and a 48-bit frame counter that can be related to the GPS timestamp
to tag every frame group with an absolute time. Each packet is also followed
by a status word indicating whether packets were lost during the various phases
of the corner-turn. The packet format is illustrated in Figure~\ref{Fig:packet_format}.

The data transiting between the corner-turn modules inside the FPGA travels as
32-bit Advanced eXtensible Interface (AXI)-Streaming buses clocked at 240~MHz.
The data links that leave the motherboard are serialized using the FPGA
multi-gigabit transceivers capable of 10~Gbit/s in both directions. These transceivers have many features controllable by software
such as transmitter power and receiver line equalizers, as well as providing
built-in logic for bit error rate measurements and live eye-diagram estimation.

The data transferred with the transceivers between FPGAs uses a stripped down
version of the 10G Ethernet protocol \hilight{designed for this application} to
minimize logic resource utilization in the FPGA. The protocol implements the
basic start of frame, end-of frame and idle codes in addition to a 32-bit cyclic
redundancy check word that validates the integrity of the payload.
Though the transmission has been tested and validated at 10~Gbit/s,
we choose to
\hilight{under-clock these links to operate at a line rate of }7.58~Gbit/s. This minimizes the FPGA internal clock rates \hilight{and} power draw, and maximizes signal integrity.

The links to the GPU are implemented with industry-standard 10G Ethernet and
send the payload using UDP packets. Again, a custom transmit-only 10G Ethernet
module was designed to provide the smallest logic footprint.

\begin{figure}[!htbp]
  \centering
    \includegraphics[width=\textwidth]{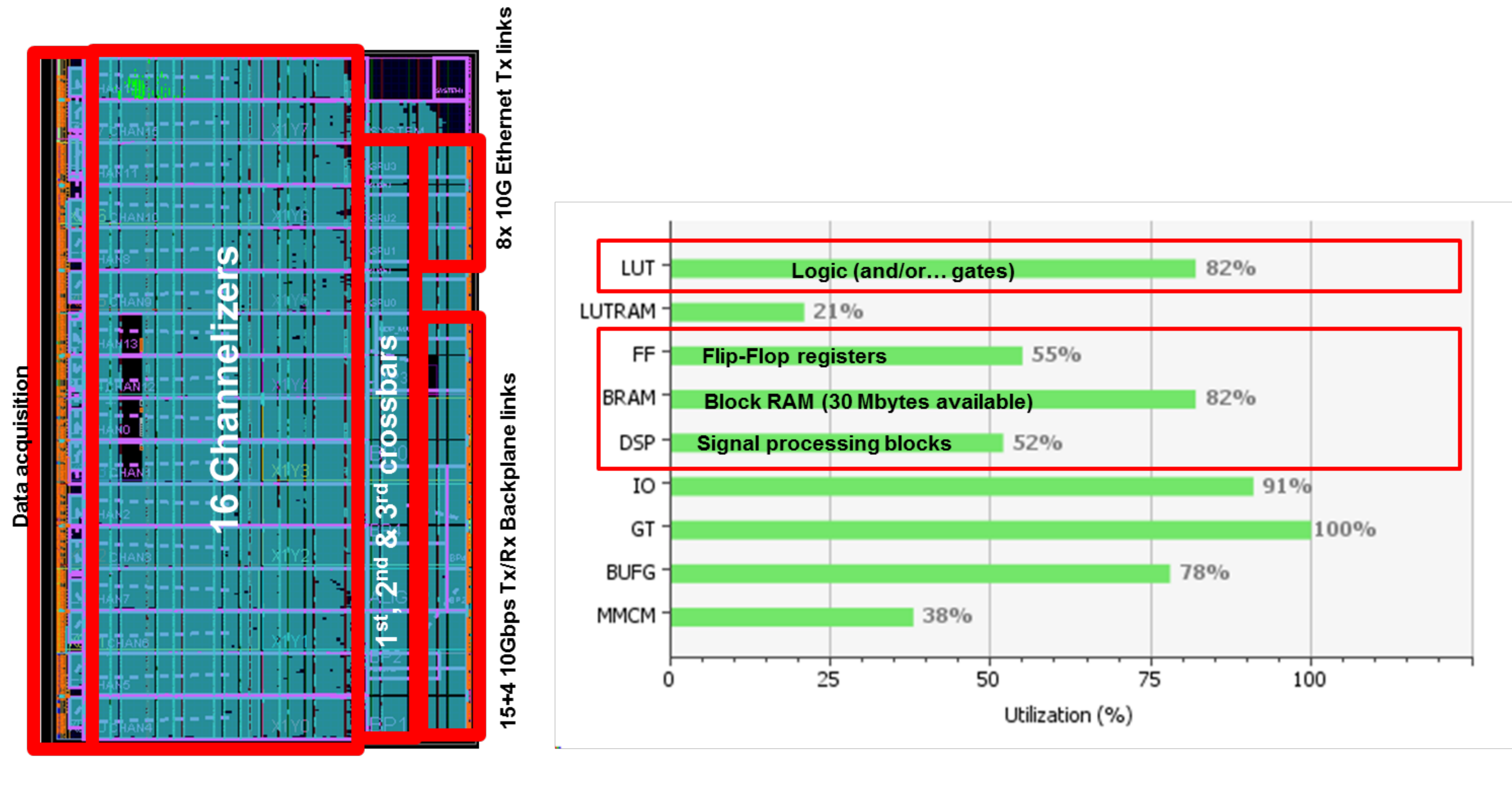}
    \caption{Resources used on the on the Xilinx FPGA to implement the F-Engine and the corner-turn logic. 
    }
  \label{Fig:fpga_resource_utilization}
\end{figure}

The corner-turn firmware building blocks described above can be parametrized,
scaled and combined to accommodate various corner-turn architectures. Figure
\ref{Fig:fpga_resource_utilization} illustrates the layout and amount of logic
used by both the CHIME F-Engine and corner-turn logic for the CHIME application.
The overall design uses more than 82\% of the embedded memory blocks, 52\% of
signal processing blocks, and 82\% of the look-up-tables within the Kintex-7
FPGA. The design takes less than
two hours to compile using the latest Vivado tools from
Xilinx. The corner logic uses less than 20\% of the FPGA area.

 %

%--------------------------------------------------------------
%!TEX root = ICE-CornerTurn_main.tex

\section{Performance}
\label{s_performance}

The performance of the CHIME's corner-turn system has been tested and validated
\hilight{both in the laboratory and in the field}.

\begin{figure}[htbp]
\centering
\includegraphics[width=.9\textwidth]{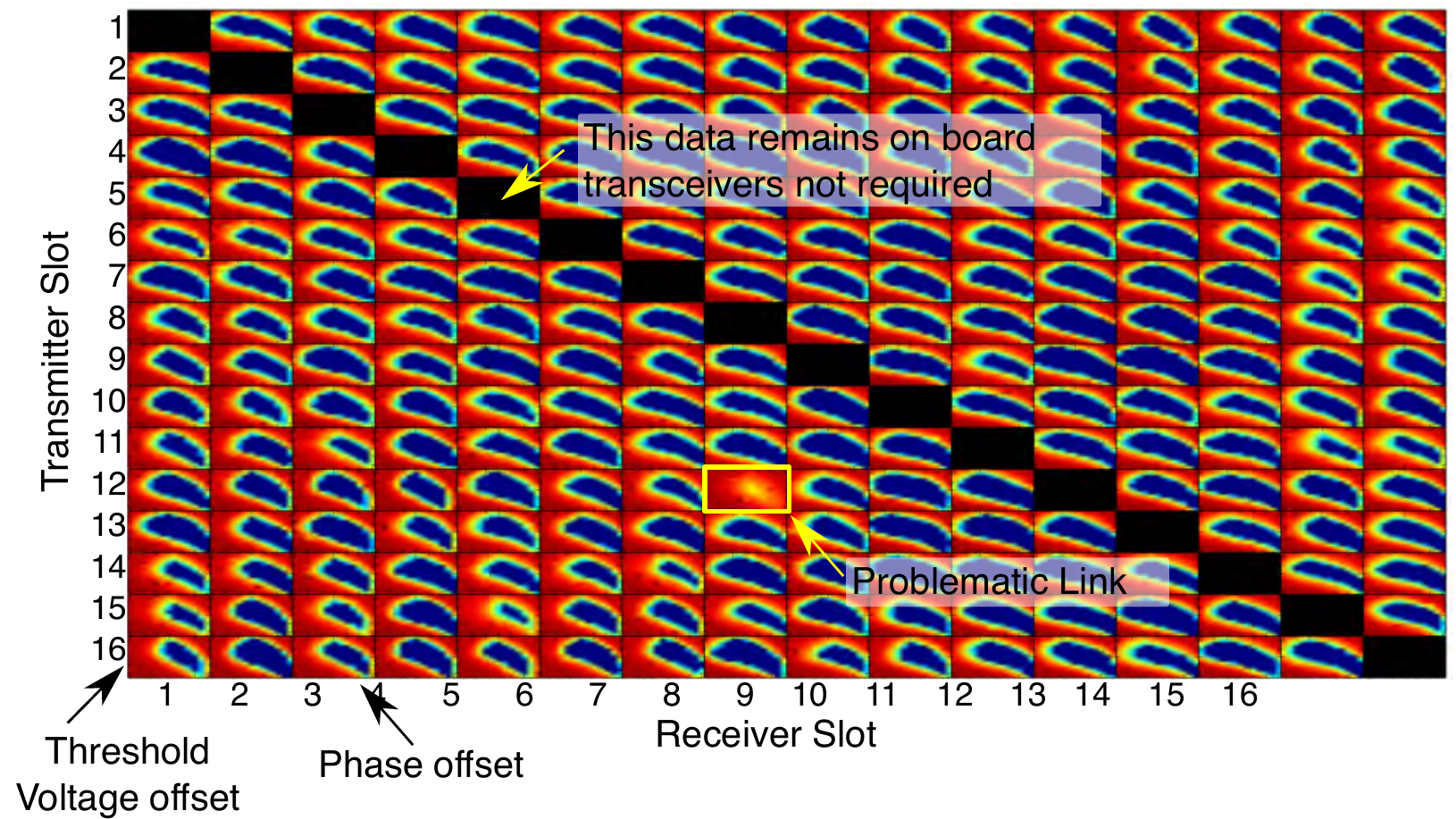}
\caption{Eye diagram of the backplane full-mesh links between all 16 motherboards in a crate at 10~Gbit/s. The diagonal has no data because the link between a board and itself is done internally. One defective link at transmit slot 12, receive slot 8 was revealed
by this particular test. The poor performing link was 
caused by a motherboard connector assembly issue that was easily identified in this manner and fixed.
 }
\label{FigFullMeshCrate}
\end{figure}

A first level of verification of the backplane full-mesh network signal integrity
 was performed by measuring the \hilight{eye diagram}\hilight{, a triggered, repeated analog measurement of the digital signal} of every link \hilight{at 10~Gbit/s}. This was
achieved by transmitting a Pseudo-Random Binary
Sequence (PRBS) between every motherboard. The \hilight{eye diagrams} were then estimated
by using the functionality built into the FPGA's 10~Gbit/s transceivers, which
operate by measuring the difference between the bits sampled at the optimum time and threshold, and the bits sampled at an offset from this optimum point. The results shown in Figure \ref{FigFullMeshCrate} reveals that all
``eyes'' \hilight{are well opened with a good sampling point}, except for one at transmit slot 12, receive slot 8. This eye diagram illustrates how a bad link can be identified. In this case, the poor performing link was 
caused by a motherboard connector assembly issue that was easily identified in this manner and fixed.

The inter-crate QSFP+ links were similarly tested by
measuring the Bit Error Rate (BER) of all the links with a PRBS sequence. 

These
tests validated both the backplane full-mesh links and the backplane QSFP+ links.
It enabled the transmit power to be adjusted to the optimal level required to
obtain error-free transmission during the test duration (many seconds), which
implies a BER lower than $10^{-11}$.

Finally, a full two-crate system feeding data to the four 10G Ethernet ports of
single GPU node was tested. The firmware was set-up to feed test patterns to the
channelizers instead of random analog samples. The performance of the system was
verified by counting the number of packets that were rejected because of cyclic
redundancy check (CRC)
errors, by monitoring the status flags of the corner-turn firmware (buffer
over/underflows, missing frames etc.), and by checking that the test pattern
data ended up on the right link in the right order.

No errors were detected during a \hilight{many-minute} test period on the backplane full-mesh
corner-turn network, the inter-crate QSFP+ passive copper links, and the 100~m
multimode fiber link to the GPU node's Ethernet interface card. The only
observed packet loss is believed to be caused by the internal handling of the
packets within the GPU node, quantified at about 0.01\%, which with further 
software tweaking has been measured to be zero on \hilight{many-minute} tests between an
individual ICE motherboard and single GPU node.

These tests show that the overall reliability of the corner-turn is sufficient for CHIME,
\hilight{where known losses at the percent level can be tolerated}.

A large portion of the  CHIME's corner-turn system has been exercised in the
field on the CHIME Pathfinder telescope, which has been operational for two
years. The Pathfinder is a prototype of the full CHIME system described in this
paper. It consists of two smaller cylinders providing signals to a single-crate,
256 input, 400~MHz-bandwidth F-engine and corner-turn system using the same
hardware and firmware as that of full CHIME.  For the corner-turn, the only
differences are that the stage three inter-crate corner-turn is not necessary and
the system uses short 7~m copper cables to connect to the GPU nodes.  The
Pathfinder corner-turn system has performed according to specifications and with
no reliability issues.

 %

%--------------------------------------------------------------
%!TEX root = ICE-CornerTurn_main.tex

\section{Alternative Design Comparison}
\label{s_comparisons}

The corner-turn hardware for CHIME is designed to maximize the
simplicity and reliability of the system, using minimal connectors and
cables, while keeping the overall hardware cost low and minimizing technical and schedule risks. \hilight{It is designed to make optimal use of the FPGA's high speed serial links.}

For the ICE system, the majority of the networking is implemented on the custom
backplanes, or through direct copper links between backplanes. Irrespective of the corner-turn strategy,
the design and
fabrication of a backplane would have been necessary to
provide clocking and power to the motherboards. Since the networking
architecture is implemented almost entirely as copper traces within
the backplanes, the additional cost of the networking features are
low: the cost of the high speed Molex Impact connectors
that interface the motherboards to the backplanes, the additional cost of using low loss materials for the backplane
circuit boards and increased layer count.

For existing radio correlators,
it has been a more common approach  to use commercial
switches for the corner-turn networking.  Present-day examples have
sufficiently small total bandwidth such that the corner-turn can be
contained within a single commercial switch.  The Long Wavelength
Array (LWA) correlator processes 58~MHz of bandwidth from 512
digitizers at 8~bits, for a total of 236~Gbit/s of data. It uses a
single Mellanox SX1024 \hilight{switch with 48 10~GbE ports and 12 40~GbE ports} ~\citep{2015JAI.....450003K}.
PAPER-64~\citep{2015ApJ...809...61A} processes 100~MHz of bandwidth from 128 digitizers using 32
high speed links, requiring a single 10~GbE switch.

For full CHIME, a matrix of switches would be needed.
Data is flowing in one direction from the eight 10~Gbit/s Ethernet ports on each
of the 128 ICE motherboard channelizers (a total of 1024 ports) to the
four 10~Gbit/s Ethernet ports on each of the 256 GPU X-engine nodes.  This means
the switch matrix must support ports with twice the total bandwidth of
the system, since each port is used in one direction
only.

\begin{figure}[htbp]
    \centering
    \includegraphics[width=.5\textwidth]{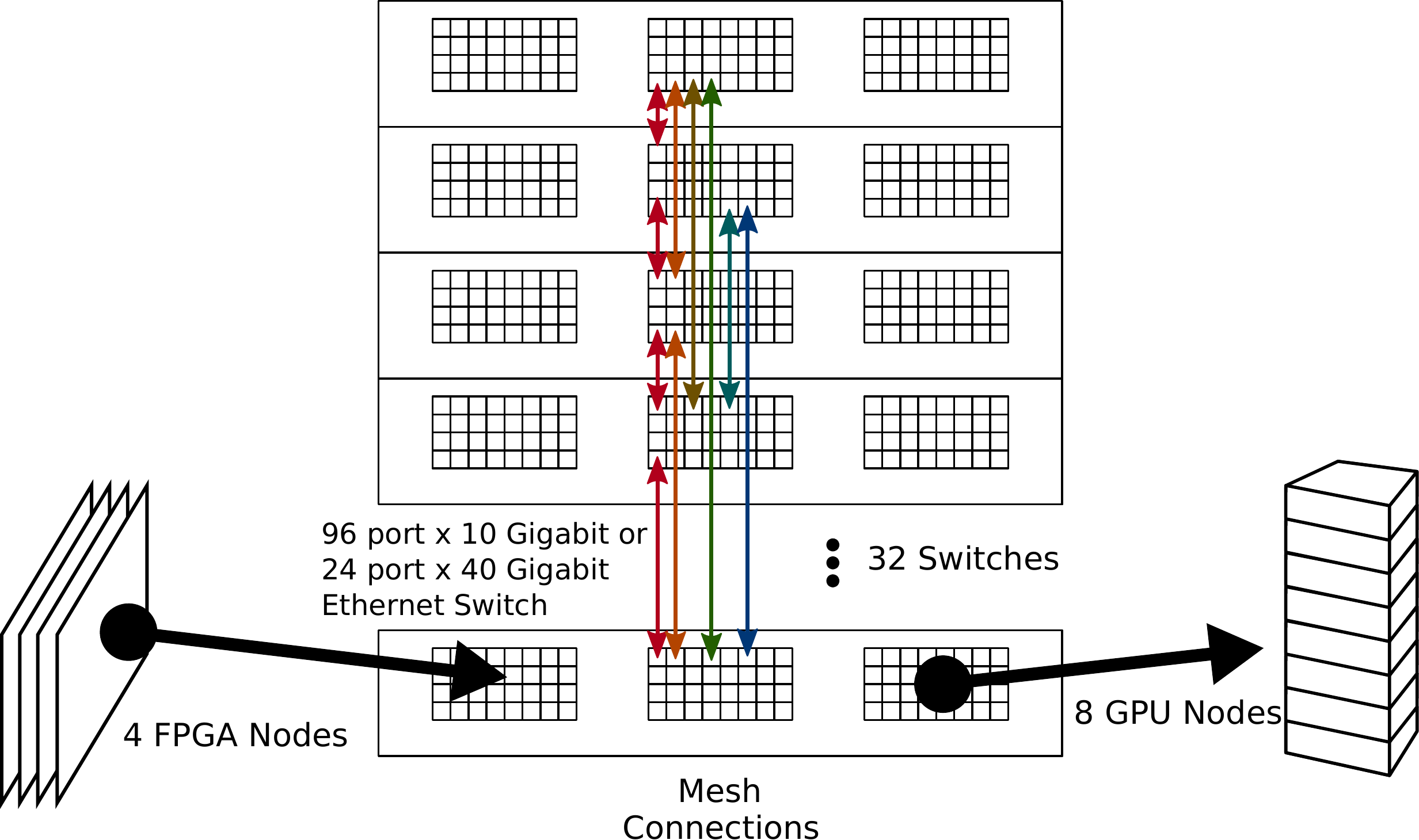}
    \caption{
    An alternative corner-turn network for CHIME could be
      constructed from a matrix of commercial switches. The CHIME
      F-engine outputs its data from 1024 10~Gbit/s Ethernet ports. An
      efficient matrix for CHIME would use $P=32$ switches, each with
      $3P=96$ ports for a total of 3072 ports in the switch matrix. One third
      of these ports connect to the F-engine, a third to the
      X-engine, and a third are used to inter-connect the 32
      switches. This commercial switch network is an order of
      magnitude more expensive than the custom solution, but more
      importantly, it poses challenges in cabling (passive cables are
      generally restricted to short, 7~m length), reliability, and
      requires the use of much smaller data packets that have large bandwidth overhead and require extra processing by the GPU nodes.
}
    \label{fig:SwitchCornerturn}
\end{figure}

An efficient configuration involves $P$ switches, each with $3P$
ports. For each switch, $P$ ports are used for input from the
F-engine, $P$ ports are used for the output to the X-engine, and
$P$ ports connect to the other switches in the matrix.
For CHIME, there would be a
total of 3072 ports for the network switch matrix, which is composed
of $P=32$ switches, each with $3P=96$ ports, as shown in
Figure~\ref{fig:SwitchCornerturn}.\footnote{This solution is not
  unique and many configuration variants exist.
  For example, in the CHIME system, since
  each destination X-engine node has four ports, it would be possible
  to divide the system into four separate networks, where
  each motherboard and each GPU node is connected to each of the four
  networks. The same number of ports and cables is required, but in
  this scenario, each of the four networks needs to interface only 256
  ports.  Thus the system can be built from 4 independent networks with
  16 switches of 48 ports each.
}
The system has 1024 cables connecting the switch matrix to the
channelizer, 1024 cables connecting the switch matrix to the X-engine, and
512 cables that are used for interconnecting the switches. At an
approximate cost of \$100 per switch port and \$25 per cable (which is
reasonable today when purchasing in volume), the total cost is
approximately \$371k, about an order of magnitude larger than the
total cost of the eight backplanes used for full CHIME. Much more
important than the cost are the other benefits provided by the custom corner-turn solution.
Cabling and maintaining
a switch matrix is a challenge. \hilight{Since 10~Gbit/s Ethernet passive cables typically have a maximum
length of less than 10~m, there are significant limitations to the physical location of each element.
The reliability of cables and cable connectors has proven much lower than the backplane interconnect, and new opportunities for human error are introduced in the cabling.}

Risk mitigation also played a major role in the decision. At the time of design, it
was not clear how a switch array would handle the massively synchronous traffic
pattern generated by the F-engine on every port of every switch. Also, the GPU
node must maximize the data transfer between the Ethernet
interface and the GPU processor memory through the node's PCI
Express bus. Using a switch-based system would have meant that a large number
of small packets would have to be reordered and processed in memory. It was
unclear whether the GPU nodes could afford those operations with the software
drivers provided by the card manufacturers, and it was beneficial to reserve the GPU node capabilities for other tasks. The custom corner-turn allowed the
FPGA to repackage the data from each corner-turn stage into large contiguous
blocks that can be easily processed by the GPU node.

A benefit of our custom network is that the FPGA sends and receives
the data multiple times, allowing it to re-packetize the information
into large, easy to parse, sequential packets for the GPU nodes to
receive, un-pack, and process. A system that uses traditional
networking switches would be forced to send either very small packets
from the individual boards in the channelizer, or maintain very large
memory pipelines in the channelizer. The packets could also arrive in any
order. This substantially complicates
and adds risk to the development of the data handling in the GPU
nodes, and requires extra memory in the channelizer hardware. Another risk with switches is that they are allowed to drop the UDP packets for any reason. With the custom networking
hardware, every packet transmission is synchronous and deterministic. The system could be tested end to end in the prototype phase
with early revision circuit boards, then scaled up to the full CHIME
implementation with very little risk and little additional engineering
time.

These factors motivated our decision to accept the small loss of the redundancy
and software-reconfigurability offered by the switch in favor of the cheaper,
lower-risk switch-free solution.

We note that, since the custom ICE corner-turn system uses standard Ethernet packets between F- and X-engine, the potential for scaling up to much larger systems exists by using 10~Gbit/s Ethernet switches between the ICE F-engine and GPU X-engine, in addition to the corner-turn described above. Switches could be inserted between the GPU nodes and optical fibers. This hybrid system would have the benefit that the FPGAs have already assembled the data into large packets before they are fed into the switches.

%--------------------------------------------------------------
\section{Conclusions}
\label{s_conclusions}

A custom full mesh corner-turn network for radio
astronomy applications has been developed and applied to the CHIME instrument. The implementation is comprised of direct FPGA to FPGA
connections within a crate of sixteen ICE motherboards, similar links between crates of boards, and direct connections from each ICE motherboard
to GPU X-engine nodes.  This allows for simple communication protocols, error handling, and data alignment all within the FPGA, and offers the potential to scale to much larger applications. The system scales from sixteen antenna inputs to the 2048 inputs required by CHIME, and far beyond. Large correlators may be implemented by increasing the number of crates in each quadrant from two to three or even five with full hardware support present.

The corner-turn solution has been implemented on the CHIME
pathfinder and is found to be very reliable, as expected
from a passive network, with error free operation between F-engine and X-engine. Combined with lab tests to ensure compatibility with the $N$=2048 input, 400 MHz configuration, the system has been validated from end-to-end for the 6.6~Tbit/s implementation for full CHIME.

The full CHIME FX correlator is scheduled for commissioning in late
2016.

%--------------------------------------------------------------
\section{Acknowledgments}
\label{s_acknowledgments}

We are very grateful for the warm reception and skillful help we have received from the staff of the Dominion
Radio Astrophysical Observatory, operated by the National Research Council Canada.

We acknowledge funding from the Natural Sciences and Engineering Research Council of Canada, Canadian Institute for Advanced Research, Canadian Foundation for Innovation and le Cofinancement gouvernement du
Qu\'ebec-FCI.
We acknowledge generous support from the Xilinx University Program.

%--------------------------------------------------------------
%%%%% References %%%%%
%%\section*{References}

%%% I use: http://scieng.library.ubc.ca/coden/
%%% for Journal Title Abbreviations

%\bibliographystyle{plain}
\bibliographystyle{ws-jai}
\bibliography{CornerTurnReferences}   % bibliography data

\end{document}